\documentclass[twocolumn,showpacs,preprintnumbers,amsmath,amssymb]{revtex4}

\usepackage{graphicx}% Include figure files
\usepackage{dcolumn}% Align table columns on decimal point
\usepackage{bm}% bold math
\newcommand{\bra}[1]{\langle #1|}
\newcommand{\ket}[1]{|#1\rangle}

\newcommand{\kdd}[0]{\ket{\!\downarrow\downarrow}}
\newcommand{\kdu}[0]{\ket{\!\downarrow\uparrow}}
\newcommand{\kud}[0]{\ket{\!\uparrow\downarrow}}
\newcommand{\kuu}[0]{\ket{\!\uparrow\uparrow}}
\newcommand{\bdd}[0]{\bra{\downarrow\downarrow\!\!}}
\newcommand{\bdu}[0]{\bra{\downarrow\uparrow\!\!}}
\newcommand{\bud}[0]{\bra{\uparrow\downarrow\!\!}}
\newcommand{\buu}[0]{\bra{\uparrow\uparrow\!\!}}

\newcommand{\du}[0]{\downarrow\uparrow}
\newcommand{\ud}[0]{\uparrow\downarrow}

\newcommand{\dm}[0]{|\!\!\downarrow\rangle}
\newcommand{\um}[0]{|\!\!\uparrow\rangle}
\newcommand{\da}[0]{|^1S_0\rangle}
\newcommand{\ua}[0]{|^3P_1\rangle}
\newcommand{\D}[2]{\ket{D^{{\scriptscriptstyle\left(#2\right)}}_{{\scriptscriptstyle#1}}}}
\newcommand{\BD}[2]{\bra{D^{{\scriptscriptstyle\left(#2\right)}}_{{\scriptscriptstyle#1}}}}

\begin{document}

\title{Preparation of Dicke States in an Ion Chain}

\author{D. B. Hume}
\email[]{david.hume@boulder.nist.gov}
\author{C. W. Chou}
\author{T. Rosenband}
\author{D. J. Wineland}

\affiliation{Time and Frequency Division, National Institute of Standards and Technology, Boulder, Colorado 80305}

\begin{abstract}
We have investigated theoretically and experimentally a method for preparing Dicke states in trapped atomic ions. We consider a linear chain of $N$ ion qubits that is prepared in a particular Fock state of motion, $|m\rangle$. The $m$ phonons are removed by applying a laser pulse globally to the $N$ qubits, and converting the motional excitation to $m$ flipped spins. The global nature of this pulse ensures that the $m$ flipped spins are shared by all the target ions in a state that is a close approximation to the Dicke state $\D{N}{m}$. We calculate numerically the fidelity limits of the protocol and find small deviations from the ideal state for $m = 1$ and $m = 2$.  We have demonstrated the basic features of this protocol by preparing the state $\D{2}{1}$ in two $^{25}$Mg$^+$ target ions trapped simultaneously with an $^{27}$Al$^+$ ancillary ion.
\end{abstract}


\maketitle

%\section{Introduction   \label{intro}}
Entanglement is a fundamentally nonclassical feature of quantum mechanics and has been recognized as an important resource in quantum information science \cite{Nielsen2000}. The coherent manipulation of entangled quantum states can be useful for important tasks such as computation, cryptography, and metrology.  In addition, the insights gained from research with entangled states may contribute to a better understanding of the foundations of quantum theory and shed light on physics at the boundary between the quantum and classical worlds.

%Various entangled states have been experimentally demonstrated. Bell states have been prepared in many bipartite quantum systems such as photon pairs \cite{Kok2007}, trapped ions \cite{Blatt2008}, and solid-state systems \cite{Korotkov2009}. For systems with multiple parts, GHZ states \cite{Leibfried2005, Sackett2000}, W states \cite{Haffner2005}, and cluster states \cite{Bloch2008} have been realized. As the number of parts in a quantum system grows, one can find more classes of entangled states \cite{Dur2000} and potentially more interesting applications, but the state preparation also becomes more challenging. Here we explore numerically a scalable approach to preparing multiple trapped ions in a general class of entangled states called Dicke states.  We then demonstrate our protocol on a mixed-species ion chain composed of two Mg$^+$ ions coupled to a single Al$^+$ ion.

Various entangled states have been experimentally demonstrated in a wide variety of physical systems including photons, condensed matter systems, atoms in optical lattices, and trapped ions \cite{Nature2008}. In the case of trapped ions, Bell states have been generated with high fidelities \cite{Leibfried2003, Roos2008}.  For larger numbers of ion qubits, GHZ states \cite{Leibfried2005, Sackett2000} and W states \cite{Haffner2005} have been realized. As the number of physical parts in a quantum system grows, we can find more classes of entangled states \cite{Dur2000} and potentially more interesting applications, but the state preparation and verification becomes more challenging. Here we explore numerically an approach to preparing multiple trapped ions in a general class of entangled states called Dicke states.  We then demonstrate the basic  protocol on a mixed-species ion chain composed of two Mg$^+$ ions coupled to a single Al$^+$ ion.

The Dicke state $\D{N}{m}$ is the equal superposition of all basis states of $N$ qubits having exactly $m$ excitations \cite{Dicke1954}. If we denote the qubit states as $|\!\!\downarrow\rangle$ and $|\!\!\uparrow\rangle$, we can write an arbitrary Dicke state in the following form
\begin{equation}
\D{N}{m}
=\binom{N}{m}^{-\frac12}\displaystyle\sum_{k}P_k\left(|\!\downarrow^{\otimes(N-m)}\uparrow^{\otimes m}\rangle\right).\label{eq:Dicke}
\end{equation}
The sum is over all $\binom{N}{m}$ permutations (produced by the permutation operator $P_k$) with $m$ qubits in the state $|\!\!\uparrow\rangle$. The W states are a special case of $\D{N}{m}$ with $m=1$. Entanglement in Dicke states is highly resilient against external perturbations and measurements on individual qubits \cite{Stockton2003, Dur2001}. Through projective measurements on some of the qubits in the system, we can obtain states of different entanglement classes. Thus, the Dicke states can serve as a versatile resource for the preparation of multipartite entangled states.

In addition to the technique used in \cite{Haffner2005}, other methods for the generation of Dicke states with trapped-ion qubits have been proposed \cite{Retzker2007, Linington20082}.  The basic features of our approach follow those proposals.  We consider a chain of $N$ ion qubits each initialized in the state $\ket{\!\downarrow}$ and collectively cooled to the ground state of motion for a particular mode.  By addressing a single ion in the chain with a laser pulse tuned to the $m^{th}$ higher-frequency motional sideband (blue sideband) \cite{Meekhof1996} we produce the state
 \begin{equation}
|\psi\rangle = \ket{\!\downarrow_1\downarrow_2\ldots\downarrow_{\rm{N}}} \otimes |m\rangle_{\rm{M}}, \label{eq:StartState}
\end{equation}
where we have labeled the motional Fock state with a subscript $\rm{M}$.  From $\ket{\psi}$, a laser pulse of appropriate duration addressing all $N$ qubits and tuned to the first lower-frequency motional sideband (red sideband), creates a state that is a close approximation of $\D{N}{m}$.

We define the fidelity for an arbitrary final state, $\rho$, as $F \equiv \BD{N}{m}\rho\D{N}{m}$.  In the case with $m=1$ (W states), the procedure outlined above can achieve arbitrarily high fidelity.  For higher-order Dicke states attainable fidelity is reduced because the red sideband pulse does not transfer population completely from $\ket{\psi}$ to $\D{N}{m}$.  The previous proposals showed that this imperfection can be mitigated by postselection \cite{Retzker2007}, or by a generalization of adiabatic rapid passage \cite{Linington20081, Linington20082}.  However, the simplified method presented here is sufficient to achieve fidelities limited by other experimental imperfection for a range of Dicke states.

One experimental challenge in the above scheme is to individually address a single ion to produce the state $\ket{\psi}$.  High confinement frequencies, and consequently small inter-ion spacing, are desirable to resolve the motional sideband spectrum.  This makes individual spatial addressing difficult.  We avoid this problem by introducing to the $N$-qubit ion chain a single ancillary ion of a different atomic species.  Simultaneous trapping of individual ions of two species has been used for sympathetic cooling \cite{Barrett2003} and indirect state detection \cite{Hume2007}.  Spectroscopic resolution of the two atomic species ensures that any laser pulse applied to one species will leave the internal states of the other ion species unchanged.  The state $\ket{\psi}$ can be created by applying a global laser pulse tuned to the $m^{th}$ sideband (or equivalently $m$ pulses sequentially tuned to the first sideband) of an accessible transition in the ancilla.

\begin{figure}
\includegraphics[scale = .6, clip=true, viewport=0in 1.1in 5.7in 7.2in]{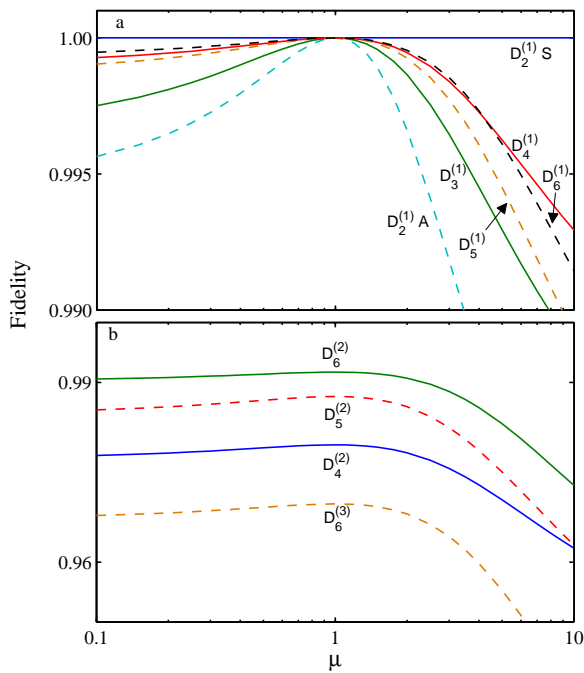}
\caption{(a) Fidelity of $\D{N}{1}$ state generation (by use of the in-phase axial mode of motion) as a function of mass ratio, $\mu$, for $N = 2\ldots6$.  In the case of $N = 2$ we consider both the symmetric case labeled S, with the ancillary ion in the center, and the asymmetric case labeled A, with the ancillary ion on the outside.  The latter case is the one chosen in the experiment.  For $N = 3$ to $N = 6$ we assume that the ancillary ion occupies a central position in the ion chain. (b) Fidelity of $\D{N}{2}$ and $\D{N}{3}$ state generation as a function of mass ratio for $N = 4\ldots6$. Here the ancilla ion occupies a central position for all cases.}\label{fig:WFid}
\end{figure}

Two-species ion chains typically contain ions of unequal mass, which leads to unequal motional mode couplings.  To see how this affects state fidelity, we consider the generation of W states ($\D{N}{1}$), for which the fidelity can be perfect in the case of equal masses.  For the $i^{th}$ ion in the chain, we denote the ground-state motional amplitude as $z_i$.  The Lamb-Dicke parameter is defined by $\eta_i \equiv k z_i$, where $k$ is the laser beam $k$-vector projection along the $\hat{z}$ direction.  In the Lamb-Dicke limit ($\eta_i \ll 1$), the strength of the ion's coupling to the red sideband pulse is $\Omega_i = \Omega_0\eta_i$, where $\Omega_0$ is the carrier ($\dm\ket{m}_M \rightarrow \um\ket{m}_M$) Rabi rate. Beginning from the state $|\psi\rangle$ with $m = 1$, under the red sideband interaction, the $N$-ion state acts like a two-level system as described by the Morris-Shore transformation \cite{Morris1983}.  The motion undergoes Rabi oscillations between the two Fock states $|1\rangle$ and $|0\rangle$ at a frequency, $\Omega^\prime$ that obeys the relation
\begin{equation}
\Omega^{\prime 2} = \displaystyle\sum_{i=1}^{N}\Omega_i^2
\end{equation}
with all $\Omega_i$ real.  Meanwhile the qubits evolve from $\ket{\psi}$ to the state
\begin{equation}
|\psi^\prime\rangle = \frac1{\Omega^\prime}\Big(\Omega_1 |\!\uparrow\downarrow\ldots\downarrow\rangle + \Omega_2 |\!\downarrow\uparrow\ldots\downarrow\rangle+\ldots+\Omega_N|\!\downarrow\downarrow\ldots\uparrow\rangle\Big),
\end{equation}
in which the terms in the superposition are the same as those in $\D{N}{1}$ but the state amplitudes are proportional to the individual ion coupling strengths, $\Omega_i$. The obtainable W-state fidelity, $F = |\BD{N}{1}\psi^\prime\rangle|^2$, can be written in terms of the coupling strengths as
\begin{equation}
F = \frac{1}{N \Omega^{\prime\,2}}\Bigg(\displaystyle\sum_{i=1}^N\Omega_i\Bigg)^2.\label{Eq:FAnalytical}
\end{equation}

To quantify the effect on fidelity of different mass ratios and different ion configurations, we first calculate the motional amplitudes \cite{Kielpinski2000}.  Each configuration is specified by the number of qubit ions, $N$, and the mass ratio $\mu = M_{\rm{ancilla}}/M_{\rm{qubit}}$.  With one exception noted below we assume that the position of the ancilla is at the center of the chain (for $N$ even) or adjacent to the center of the chain (for $N$ odd).  These configurations can be prepared deterministically by adjusting trap parameters.  We calculate equilibrium positions for the $N+1$ ions, then determine the amplitude of small oscillations about equilibrium.  The in-phase, axial mode of motion exhibits the smallest deviations between individual ion motional amplitudes and is used for all fidelity calculations.  These amplitudes give us the coupling strengths, $\Omega_i$, and $F$ follows from Eq. (\ref{Eq:FAnalytical}).  The results are presented in Fig. \ref{fig:WFid}(a), where we have assumed the Lamb-Dicke limit for simplicity, although fidelities for $\{\eta_i\} \approx 1$ are similar.  The case labeled $D^{\scriptscriptstyle{(1)}}_{\scriptscriptstyle{2}}\,\, \rm{S}$ is the only one that yields a theoretically perfect fidelity for any $\mu \neq 1$ because that symmetric configuration, with the ancilla in the center of two qubits, gives equal motional amplitudes for the two outer ions.  The cases with $N = 3\ldots6$, as well as that labeled ${D^{\scriptscriptstyle{(1)}}_{\scriptscriptstyle{2}}\,\, \rm{A}}$, with the ancilla at the outside position, allow high fidelity over a wide range of mass ratios.

In Fig. \ref{fig:WFid} (b) we present the result of a similar calculation for the states $\D{N}{2}$ and $\D{N}{3}$ with $N = 4\ldots6$. During the application of the red sideband pulse the ions state evolves in a complicated manner.  To determine a practical upper limit on the fidelity, we numerically find the first maximum of the quantity $F = \BD{N}{2}\rho(t)\D{N}{2}$ as a function of the red sideband pulse duration.  Here $\rho(t)$ is the reduced density matrix of the qubit system at pulse duration $t$ after tracing over the motional degrees of freedom.  We find fidelities for the case $m=2$ as high as 0.99 and for $m = 3$ above 0.96.  The effect of unequal masses is found to be similar to the W state case with a mass ratio of $10$ reducing the optimum fidelity at the 1 \% to 2 \% level.  These numbers refer to the fidelities at the first maximum during the red sideband evolution, but, in the absence of decoherence, $F$ could in general be made higher by evolving for a longer duration and reaching a later maxima.

 Note that the fidelities for $\mu = 1$ in Fig. \ref{fig:WFid} are equal to the fidelities obtainable in the case where there is no ancilla and $\ket{\psi}$ is generated by spatially addressing a single ion with a strongly focussed laser beam.  Here, there are only $N$ ions in the chain, and they share equal motional amplitudes for the in-phase axial mode.  Under the red sideband interaction beginning from $\ket{\psi}$ the system can be transformed to a basis where it evolves as a ladder of $m+1$ states spaced equally in energy \cite{Rangelov2006}.  Here the red sideband Hamiltonian is symmetric with respect to ion exchange, so all ions participate equally in the entangled state.

%\section{Experiment}
We have demonstrated the basic features of the protocol in an experiment with one $^{27}$Al$^+$ ancillary ion and two $^{25}$Mg$^+$ qubit ions. The ions are trapped in a linear RF Paul trap \cite{Rowe2002}, with trap frequencies for a single $^{25}$Mg$^+$ ion $\{\omega_x,\omega_y,\omega_z\}=2\pi\times\{5.54,6.46,2.55\}$ MHz where z denotes the axis of the ion chain. In $^{25}$Mg$^+$ we use two Zeeman sub-levels of the ground-state hyperfine manifold as a qubit. We define  ${\dm\equiv|^2\text{S}_{\frac{1}{2}}, F=3, m_F=-3\rangle}$ and ${\um\equiv|^2\text{S}_{\frac{1}{2}}, F=2, m_F=-2\rangle}$.  In $^{27}$Al$^+$ the relevant qubit levels are defined as ${\da\equiv|^1\text{S}_{0}, F=\frac52, m_F=-\frac52\rangle}$ and ${\ua\equiv|^3\text{P}_{1}, F=\frac72, m_F=-\frac72\rangle}$. The ions are loaded into the trap via photo-ionization.

The order of the ions is maintained as {Mg-Mg-Al} by monitoring the motional spectrum and adjusting DC voltages of the trap electrodes to regain the correct order when necessary.  Specifically, we raise the DC endcap voltages and apply a radial bias field to configure a radially oriented, linear ion chain with Al at one end.  Then we apply differential endcap voltages to twist the radial chain to the desired orientation and finally relax the voltages back to the experimental parameters.  In the order {Mg-Mg-Al}, the amplitudes of motion for the Mg$^+$ ion in the in-phase mode are equal to within $1\%$, having a negligible impact on fidelity, while the Mg$^+$ ion spacing is small ($3\; \rm{\mu m}$) to facilitate equally strong interaction with the laser beams.

Initially, the three axial motional modes of the ions are cooled close to the ground state by resolved sideband Raman cooling of Mg$^{+}$ \cite{Monroe1995}.  This process fills the largest part of our experimental duty cycle, about 2 ms.  We observe residual phonon numbers $\overline{n} < 0.1$ for all axial modes.   Optical pumping ideally prepares the system in ${\kdd\da|0\rangle_M}$. The preparation of $\D{2}{1}$ starts with a laser pulse that adds one phonon to the in-phase axial mode by driving the blue sideband transition ${\da|0\rangle_{M}\rightarrow\ket{^3P_1}|1\rangle_{M}}$ in the Al$^+$ ion, changing the ion state to $\kdd\ua|1\rangle_{M}$. Then a laser pulse removes one phonon from the in-phase mode by driving the red sideband transition in the Mg$^+$ ions. This ideally produces the state ${\frac{1}{\sqrt{2}}(\kdu + \kud)\ua\ket{0}_M}$ \cite{King1998}.

%\section{Data Analysis}
\begin{figure}
\includegraphics[scale = .6,clip=true, viewport=0in .5in 5in 6in]{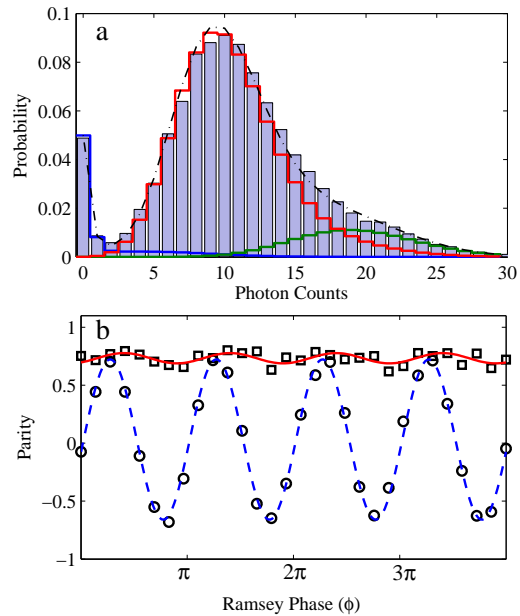}
\caption{(a) Example experimental histogram of photon counts from fluorescence detection of entangled ions (bars) and its fit (dot-dashed line). The stepped lines represent calibrated histograms $P(n|0)$, $P(n|1)$, and $P(n|2)$, scaled respectively by the fit parameters $\{c_0, c_1, c_2\} = \{0.08, 0.80, 0.12\}$. The fit gives a parity of $\Pi(0,0) =$ -0.80(1). (b) Parity of qubit state after applying a rotation $R(\frac{\pi}2,\phi)$ of varying phase (squares) and the parity after first applying the rotation $R(\frac{\pi}2,\frac{\pi}2)$, then applying a second rotation, $R(\frac{\pi}2,\phi)$, of varying phase (circles).  The residual oscillation in the former case (squares) arises from small even-parity populations in $\rho$. The measurements give a an overall fidelity $F =$ 0.77(2).}\label{fig:FData}
\end{figure}

 In terms of the individual density matrix elements, the fidelity of the final Mg$^+$ state, $\rho$, is
 \begin{equation}
 F = \frac12(\rho_{\du,\du} +\rho_{\ud,\ud} + \rho_{\du,\ud} + \rho_{\ud,\du}).
 \end{equation}
 We measure the fidelity with the same technique used in previous experiments \cite{Sackett2000}.  The odd-parity population, $\rho_{\du,\du} +\rho_{\ud,\ud}$, is measured using resonance fluorescence histograms from $\rho$.  The off-diagonal term, $\rho_{\du,\ud} + \rho_{\ud,\du}$, is inferred from the parities of the states obtained by applying a qubit rotation $R(\frac{\pi}{2},\phi)$ of variable phase. Here we use the convention that $R(\theta,\phi)$ implements the transformation, ${{\dm \rightarrow {\rm cos}\!\left(\frac{\theta}{2}\right)\dm - i e^{-i\phi}{\rm sin}\!\left(\frac{\theta}{2}\right)\um}}$ and ${\um \rightarrow - i e^{+i\phi}\rm{sin}\!\left(\frac{\theta}{2}\right)\dm + \rm{cos}\!\left(\frac{\theta}{2}\right)\um}$.  In terms of the parity operator for two qubits,
\begin{equation}
\Pi = \kdd\bdd + \kuu\buu - \big(\kdu\bdu + \kud\bud\big),
\end{equation}
we define $\Pi(\theta,\phi) \equiv tr\left(R^{\dagger}(\theta,\phi)\rho R(\theta,\phi)\Pi\right)$.  We can then write $\rho_{\du,\ud} + \rho_{\ud,\du} =
\frac12[\Pi(\frac\pi2,0)+\Pi(\frac\pi2,\frac\pi2)]$.  A signature of the state $\D{2}{1}$ is that it produces an even parity state independent of the $\pi/2$ analysis pulse phase.

We detect the state of the Mg$^+$ qubit by applying $\sigma -$polarized laser pulse of duration 200 $\mu s$ resonant with the $\dm \rightarrow |^2\text{P}_{\frac{3}{2}}, F=4, m_F=-4\rangle$ cycling transition and counting photons.  We fit observed photon histograms to a weighted sum of the distributions $P(n|0)$, $P(n|1)$ and $P(n|2)$, which correspond to the probability of observing $n$ photons given 0, 1 or 2 ions in the state $|\!\!\downarrow\rangle$.  Assuming equal illumination and equal photon collection efficiency for the two ions, we can write the distributions above in terms of single-ion count distributions as
\begin{eqnarray}
P(n|0) = P_{BG}(n)*P_{\uparrow}(n)*P_{\uparrow}(n)\\
P(n|1) = P_{BG}(n)*P_{\uparrow}(n)*P_{\downarrow}(n)\\
P(n|2) = P_{BG}(n)*P_{\downarrow}(n)*P_{\downarrow}(n),
\end{eqnarray}
where $G(n)*H(n) \equiv \sum_{m \le n}G(n-m)H(m)$ is the discrete convolution of the distributions $G$ and $H$. Here the convolved distributions refer to photon count probabilities from background scattering ($BG$), as well as the two qubit states $|\!\!\downarrow\rangle$ and $|\!\!\uparrow\rangle$.  In the case of $P_{\uparrow}(n)$ we account for off-resonant repumping from $\um$ to $\dm$ by assuming an exponential decay rate, $\gamma$, such that if the ion began in $\um$, the state probabilities at time $t$ are given by $c_{\uparrow} = e^{-\gamma t}$ and $c_{\downarrow} = 1 - c_{\uparrow}$ \cite{Langer2006}. The photon count rates and repump rate that determine the distributions result from a simultaneous fit to two reference histograms taken just before or after the experiment.

In any particular measurement, we fit the unknown distribution
\begin{equation}
P_\rho(n) = \displaystyle\sum_{i=0}^{2} c_i P(n|i)
\end{equation}
with $\sum_i c_i = 1$ to the observed series of $N_p$ samples of photon counts, $\{n_1, n_2\ldots n_{N_p}\}$. We use a maximum likelihood method where we maximize the quantity
\begin{equation}
L = \displaystyle\prod_{j = 1}^{N_p} P_{\rho}(n_j)
\end{equation}
by adjusting the parameters $c_i$.

An example of the results of measuring photon counts from the entangled state as well as a fit to the distribution are shown in Fig. \ref{fig:FData}(a).  We measure odd-parity populations of the initial entangled state of $c_1 = -0.80(1)$. In Fig. \ref{fig:FData}(b) we display the measured parity of our state as a function of the phase of the analysis pulse (squares) as well as a sinusoidal fit.  The mean value of $\Pi(\frac{\pi}{2},0)$ and $\Pi(\frac{\pi}{2},\frac{\pi}{2})$ is 0.74(2).  These numbers together yield an overall fidelity of $F = 0.77(2)$.  To further confirm the presence of entanglement, we applied the pulse $R(\frac\pi2,0)$ to $\rho$ followed by a second analysis pulse $R(\frac\pi2,\phi)$ of varying phase (circles), again measuring the parity of the final state.  Here we observe quantum state interference in the sinusoidal parity oscillation with a period of $\pi$ and an amplitude of 0.70(3).

%There are several contributing experimental sources of infidelity.  In separate  measurements we have determined that the laser cooling of the in-phase mode of motion leaves less than 0.05 quanta in higher energy states, making this contribution to the infidelity close to $3 \%$.  The preparation of the state $\da$ could only be confirmed indirectly, by mapping information to the Mg$^+$ ions \cite{Hume2007}.  This proceeds in exactly the same way as the entangling pulses and is subject to effects from imperfect ground state cooling and other noise sources described below.  With this complication we can only constrain the error to near 10\%.  We also observe non-negligible decoherence of the Al$^+$ rabi flopping due to magnetic field noise which will contribute errors at the level of $3 \%$.  The 300 $\mu s$ lifetime of the $^3P_1$ state leads to a spontaneous emission probability of 0.05 during the 15 $\mu s$ pulse and an infidelity of approximately $3 \%$.  If the two Mg$^+$ ions .  Even at the optimum beam position, beam pointing fluctuations can cause errors, which we estimate to be close to $5 \%$.

Several experimental imperfections contribute to the overall infidelity of $1 - F = 0.23(2)$.  The largest error stems from random failures to optically pump to the Al$^+$ $|^1S_0, m_F=-\frac52\rangle$ ground state at the 10 \% level, which has been tested in separate experiments.  The next largest error is due to beam pointing fluctuations.  If the two Mg$^+$ ions are unequally illuminated by the Raman beams during the entangling pulse, two errors arise.  The state amplitudes will accumulate a differential phase due to unequal Stark shifts and the state probabilities will be different due to unequal coupling strengths.  We balance the Stark shifts by suppressing the beating of fringes in a Ramsey experiment, in which each Raman beam is applied separately during the wait period.  Even at the optimum beam position, beam pointing fluctuations can cause errors, which we estimate to be approximately 5 \%.  Additional errors are caused by imperfect ground-state cooling of the in-phase mode of motion (0.05 quanta remaining, 3 \% infidelity), Al$^+$ decoherence from laser and magnetic field noise (3 \% infidelity), and Al$^+$ spontaneous emission (300 us upper state lifetime) during the $|^1S_0\rangle|0\rangle_M \rightarrow |^3P_1\rangle|1\rangle_M$ pulse ($15\, \mu s$ pulse duration, 1 \% infidelity).  These estimated errors add to 0.22 and agree with the experimental fidelity.

In summary, we have explored numerically and experimentally a protocol for the creation of Dicke states in a trapped ion chain.  The pulse sequence we use involves just two consecutive laser pulses, tuned to the resonance of a motional sideband and addressing all ions simultaneously.  Infidelities due to unequal ion masses and inequivalent positions in the chain can be small ($\ll1 \%$) for creating W states, and the infidelity for creating Dicke states of two and three excitations can be sufficiently low to enable future interesting experiments.  We have demonstrated the basic features of the process on an ion chain composed of two Mg$^+$ ions coupled to a single Al$^+$.  If the technical errors observed in this demonstration are reduced, scaling the experiment to a larger number of ions would require the same number of steps, which makes it an attractive method for enabling the study of multipartite entangled states.

We thank Jonathan Home and Yves Colombe for helpful comments on the manuscript and acknowledge the support of IARPA, DARPA, NSA, ONR, and the NIST Quantum Information program. This Letter is a contribution of NIST and
not subject to U.S. copyright.

\end{document}